| | |
|---|---|
| Title | Ignition and extinction phenomena in helium micro hollow cathode discharges |
| Authors | M. K. Kulsreshath[1] N. Sadeghi[2] L. Schwaederle[1] T. Dufour[1] L. J. Overzet[1,3] P. Lefaucheux[1] and R. Dussart[1] |
| Affiliations | [1] GREMI, CNRS/Université d'Orléans (UMR7344), Orleans, France<br>[2] LIPhy, CNRS and Universite Joseph Fourier (UMR5588), Grenoble, France<br>[3] PSAL, UTDallas, Richardson, Texas 75080-3021, USA |
| Ref. | J. Appl. Phys., 2013, Vol. 114, 243303 (8 pp) |
| DOI | http://dx.doi.org/10.1063/1.4858418 |
| Abstract | Micro hollow cathode discharges (MHCD) were produced using 250 µm thick dielectric layer of alumina sandwiched between two nickel electrodes of 8 µm thickness. A through cavity at the center of the chip was formed by laser drilling technique. MHCD with a diameter of few hundreds of micrometers allowed us to generate direct current discharges in helium at up to atmospheric pressure. A slowly varying ramped voltage generator was used to study the ignition and the extinction periods of the microdischarges. The analysis was performed by using electrical characterisation of the V-I behaviour and the measurement of He*($^3S_1$) metastable atoms density by tunable diode laser spectroscopy. At the ignition of the microdischarges, 2 µs long current peak as high as 24mA was observed, sometimes followed by low amplitude damped oscillations. At helium pressure above 400 Torr, an oscillatory behaviour of the discharge current was observed just before the extinction of the microdischarges. The same type of instability in the extinction period at high pressure also appeared on the density of He*($^3S_1$) metastable atoms, but delayed by a few µs relative to the current oscillations. Metastable atoms thus cannot be at the origin of the generation of the observed instabilities. |

# 1. Introduction

Microdischarges can be produced by using many types of microdischarge reactors with different configurations.1 Microdischarges are spatially confined to dimensions smaller than 1mm and usually work in a non-equilibrium regime. In the last decade, many interesting properties of microdischarges have been demonstrated and studied using a variety of configurations.2,3 Atmospheric pressure microplasmas have gained interest in last few years and have some potential applications, e.g.: gas treatment technology or display technology.4,5 In the category of micro hollow cathode discharge (MHCD), there exist many different types of devices, which are working at atmospheric pressure: cathode boundary layer devices (CBL), through hole MHCD, plasma microjets, etc.4 These microdischarge devices can run in DC as well as in AC.1,2,4,6,7 But, many questions are remaining concerning their mechanisms and limitations. Optical diagnostic techniques are non invasive and can provide many informations. Moreover, they are quite suitable for the study of MHCDs.8–10 Among these optical diagnostic techniques, mainly Optical Emission Spectroscopy (OES), Tunable Diode Laser Absorption Spectroscopy (TDLAS), and Phase Resolved Optical Emission Spectroscopy (PROES) have been used to characterize microdischarges.11 Using these diagnostic techniques, the normal plasma operation of MHCDs have already been studied and presented.2,7,12,13 But only few studies are available in literature related to the study of the ignition and the collapse of the plasma inside the microcavity of a MHCD device.16,24

The aim of this paper is to study the ignition and extinction of the microplasmas obtained in an alumina based MHCD with electrical and optical characterisation techniques. For this purpose, a slowly varying (0.05 Hz) ramped voltage generator is used in order to have an almost constant applied voltage during about 100 µs duration of the observed transient phenomena in the ignition and collapse periods of the discharge. For the optical characterisation, TDLAS measurements on He*($^3S_1$) were performed. In this article, we first present the structure used for the MHCD reactors, the experimental set up and the details of the techniques used for the characterisation. Afterwards, electrical characterisations for a single hole MHCD are presented with a focus on their ignition and extinction. Then, the transient phenomena at the ignition and extinction are explored using TDLAS measurements.

## 2. MHCD structure

The MHCD reactors were built within a collaboration program between UTDallas (Plasma Science and Application Laboratory, Overzet) and the GREMI Lab. They consisted of a 250 µm thick layer of alumina dielectric ($Al_2O_3$) sandwiched between two 8 µm thick Ni electrodes layers made by an electrochemical deposition process. A through cavity at the center of the chip was formed by laser drilling technique. Drilling was performed by using Nd:YAG laser. Different diameters of MHCD were available from 250 µm to 400 µm. Note that, due to the laser process, the cavity was not perfectly cylindrical. For the experiments, MHCD devices with a cavity diameter of 300–360 µm were used, i.e., $D_1$=360 µm and $D_2$=300 µm (as shown in Figure 1).[2,7]

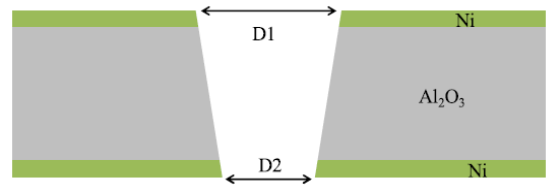

FIG. 1. Cross-section view of the structure of the MHCD reactor containing a dielectric layer of $Al_2O_3$ sandwiched between two Ni electrodes.

## 3. Experimental setup

The schematic of the experimental setup is shown in Figure 2. In these experiments, the MHCD is located inside a 2 l stainless-steel octagonal vacuum chamber in which the gas pressure could vary from $10^{-6}$ to $10^3$ Torr. The diameter and height of this vessel were 14 cm and 11 cm, respectively. On each side of the octagonal walls, circular openings were available. Three of these openings were used for the gas supply, for electrical connectors and for the sample holder positioning system. Glass windows were fitted on the other openings of the chamber for optical measurements and direct observation. The top of the vessel was covered by a stainless-steel plate and separated by a rubber gasket for the tightness. An observation glass window was also placed in the middle of the cover plate. The lower portion of the chamber was connected to the pumping system. The microdischarge reactors were installed in the chamber using a home-made sample holder. For these measurements, He gas was used. Before plasma experiments, the chamber was first evacuated down to $10^{-5}$ Torr, with primary and turbo molecular pumps. It was subsequently back filled with helium to a set pressure ranging from 100 to 1000 Torr. A Baratron gauge and a Penning gauge were used to measure the working and base pressures, respectively. A digital mass flow controller (0-100 sccm) was used to maintain a gas flow of 10-15 sccm through the chamber in order to renew the gas during the experiments.

The plasma was generated with a 0 to 2.5 kV DC (300 W) power supply. The electrical characterisations were performed by using a 200MHz oscilloscope and high voltage probes. A 50 mHz period triangular signal was sent from a function generator to the power supply to vary slowly the output DC voltage. The MHCD was connected to the power supply through a 39 kΩ ballast resistor ($R_b$). A 1 kΩ resistor was also placed between the MHCD and ground to measure the discharge current.

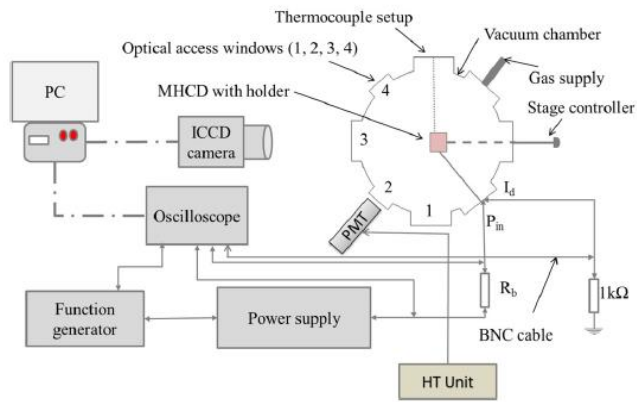
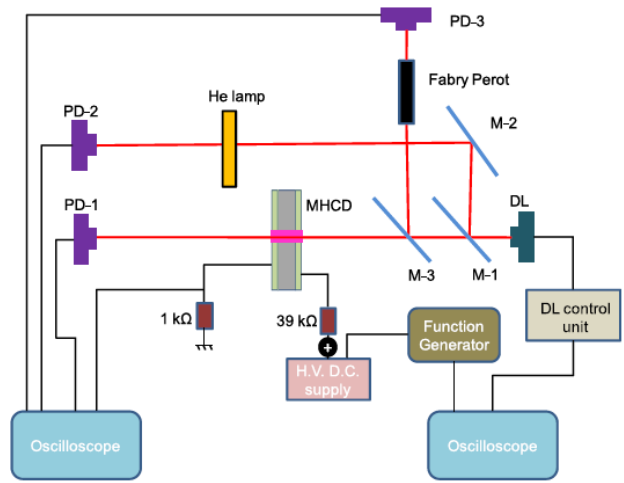

FIG. 2. Experimental setup for the DC operation of MHCD.

FIG. 3. Schematics of the experimental setup for TDLAS measurements.

The experimental setup for TDLAS is shown in Figure 3. A tunable Diode Laser (DL) emitting at 1083 nm was used for the experiments. This wavelength corresponds to the transitions between He* $^3S_1$ metastable state and $^3P_{0,1,2}$ excited states. The two components $^3P_2$ and $^3P_1$ of these lines are separated by 2.34GHz and at high temperature and/or high pressures, the linewidth of each line exceeds 5GHz. Consequently, these two lines are not resolved and the recorded absorption profile is the sum of two Voigt profiles, whose relative amplitudes are 5 and 3 (for transitions to $^3P_2$ and $^3P_1$ levels, respectively). Beam splitters were used to split the DL beam into 3 components: one of them was directed to a standard low pressure He spectral lamp, which provided a reference for the absorption spectrum profile of He at low pressure and room temperature; the second one was sent through a 20 cm long confocal Fabry Perot Etalon (FPE), whose Free Spectral Range (FSR) was 0.375GHz; and the third one was directed into the MHCD hole. Each beam was detected by an independent photodiode (PD). Signals from PDs were recorded and processed by a digital oscilloscope.

# 4. Results

## 4.1. Electrical characterisation of the ignition and extinction of the plasma

This section presents different effects observed during the plasma ignition and extinction in the MHCD reactor. A ramped DC voltage was applied to a single cavity MHCD in He gas at a particular gas pressure and V-I curves were recorded.

Figure 4 shows the standard voltage and current curves for a MHCD normal glow operating at 400 Torr He pressure. A typical electrical behaviour of the MHCD was obtained: before plasma ignition, voltage keeps on rising. At the breakdown (A), the voltage between electrodes drops and the current reaches a certain value depending on the value of the used ballast resistor. Then, the discharge voltage remains constant as far as the cathode surface is not limited, as reported in Ref. 2. Finally, with the decay of the applied voltage, the discharge current decreases until it reaches zero (B). With this time scale (in seconds), it is of course impossible to detect the transient phenomena involved at the ignition and at the extinction of the microdischarge. To study the ignition and extinction phases of the microplasma, the time scale of the oscilloscope was reduced to microsecond range. Regions (A) and (B) (Figure 4) were investigated to analyse the ignition and extinction of the MHCD.

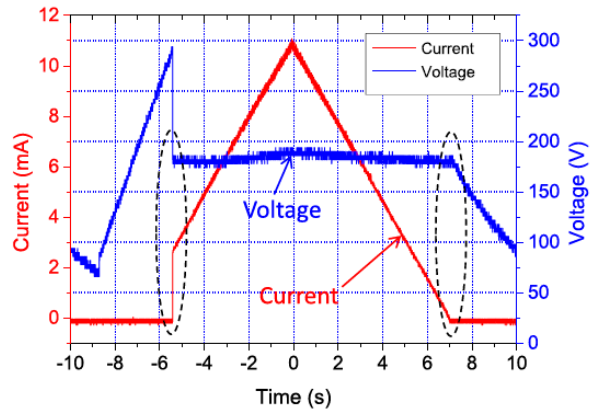

FIG. 4. Standard voltage and current plots with respect to (w.r.t.) time for a single hole MHCD reactor at 400 Torr He.

### 4.1.1. Ignition

In Figure 5, current and MHCD voltage waveforms are shown during the ignition of the plasma at (a) 350 Torr and (b) 750 Torr. The discharge inside the MHCD reactor starts with a large current pulse and reaches a steady state regime after a few oscillations. The typical pulse duration is around 2 µs FWHM (full width at half maximum). The same types of oscillations, but out of phase to the current signal, are also observed on the voltage waveform (blue dashed curve in Figure 5). In fact, before breakdown, the voltage rises slowly and charges the capacitor formed by the MHCD electrodes. Afterwards, plasma starts to ignite and a sharp voltage drop can be seen. The huge current peak results from the discharging of the capacitor through the MHCD hole and corresponds to the transient regime of the microplasma before a steady state regime is reached.12,13

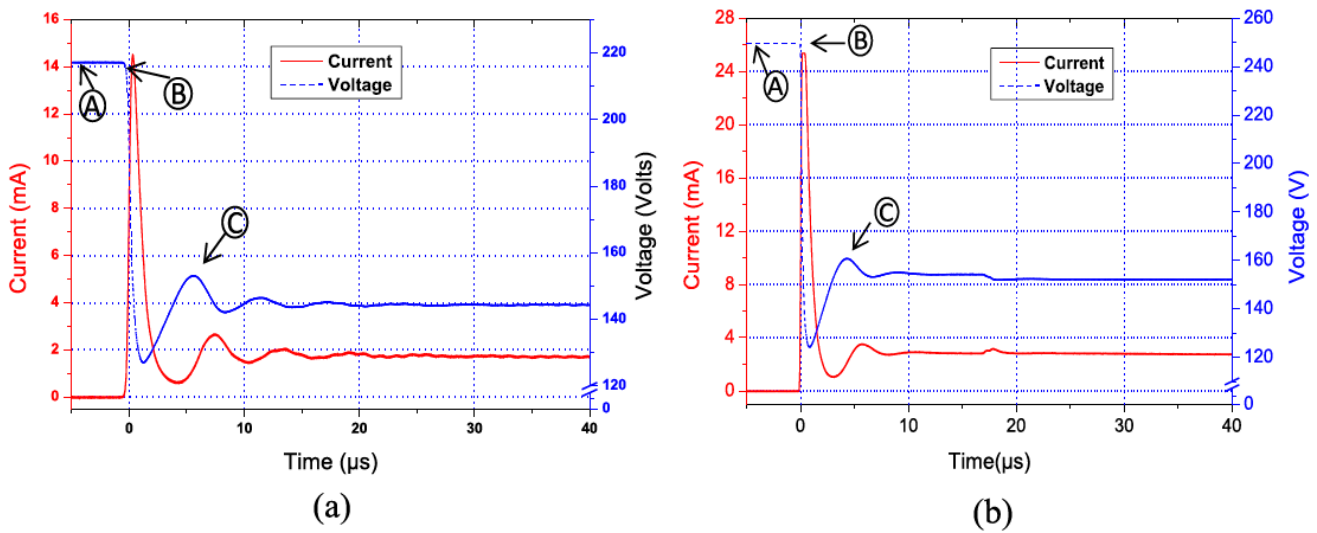

FIG. 5. MHCD current and voltage waveforms at the ignition of Helium micro discharges (a) at 350 Torr and (b) at 750 Torr.

It is possible to quantify the stabilisation time required for the microdischarges to reach the normal glow regime. The plots of Figure 5 can be explained step by step to better visualise the plots immediately after the breakdown phenomenon. We can distinguish three zones: the pre-breakdown zone (between points A and B), the transition zone (between B and C), and the normal glow (after point C onwards). In the pre-breakdown, the current is zero but the voltage increases till breakdown potential is reached. Some research teams14,15 have used the same type of geometry for microhollow cathode reactors and measured a very low current before the voltage drop and the normal glow regime, thus justifying the existence of a pre-breakdown abnormal regime. Experimentally, we cannot measure the current from our samples because our cathode thickness is about 8 µm, while this regime is observed for cathodes with a thickness of several hundreds of micrometers.

At the breakdown, the transient phenomenon appears. The current initially reaches a peak value of approximately 15mA for 350 Torr and 25mA for 750 Torr in around 300 ns and then drops and stabilises to a few mA, depending on pressure, after a few microseconds. At the microscopic level, this means that the electric field and the cathode sheath also stabilised after a few microseconds.[6,24] Moreover, this period of stabilisation depends slightly on the pressure.

The ignition signal can be compared to the self-pulsing regime.[16] This regime appears when the mean current of the micro-discharge is limited to a quite low value, typically less than 1 mA. It is quite easy with our setup to run the MHCD in the self-pulsing regime just by switching the ballast resistance to 1MΩ, which significantly reduces the mean current. Then, the current is self-pulsed as reported by several other teams.[17–19] Figure 6 shows current and voltage waveforms in this particular regime with applied voltage fixed to 380 V. At point A, the voltage increases until it reaches the breakdown potential (point B). The current appears suddenly (at point B) and reaches a value of 28mA at point C. The pulse duration is typically 1-2 µs, depending on the pressure and applied voltage.

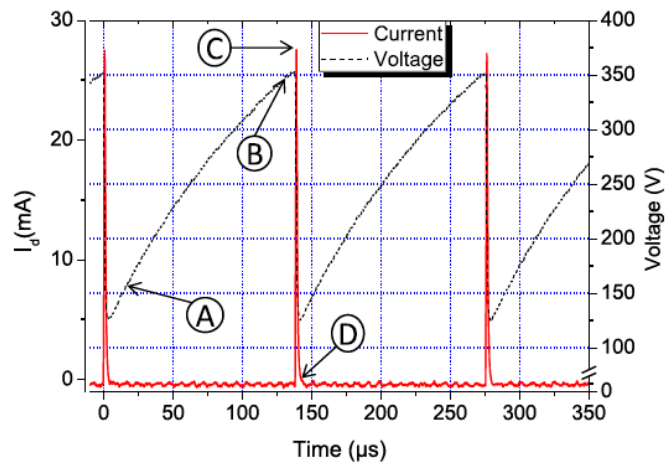

FIG. 6. Current and voltage plots vs. time for the MHCD operating in self-pulsing regime in 200 Torr helium. The ballast resistor is 1MΩ and applied voltage of 380 V.

By analogy with an RC circuit, the duration between A and B (about 140 µs) is the time needed to accumulate charges on the cathode of the equivalent capacitor of the MHCD until the voltage reaches the breakdown voltage. Here, the MHCD device capacitance is around 60 pF so that the time constant for charging, it is 60 µs for this 1MΩ ballast resistor. Then, a high current peak is obtained, which corresponds to the discharge of the capacitor through the MHCD. Between points B and C, Aubert and Rousseau showed that the discharge extends on the outer surface of the cathode.[16] Since the discharge time is much shorter than the charging time of the capacitor, the plasma switches off until the capacitor is charged again and voltage reaches again the breakdown value. As a consequence, the discharge is not self-sustained and vanishes after 2 µs (between points C and D). Actually, between these two points, if the electrodes are sufficiently thick, it was shown that the plasma is not completely off, but a weak micro-discharge can operate inside the cavity of the micro-cathode. Images obtained by ICCD camera illustrating this phenomenon are presented in the work of Aubert and Rousseau.16 Note that in our case, this weak micro-plasma is not observed because electrodes are much thinner than those used by Aubert and Rousseau. As a result, plasma cannot remain confined inside the hole, because the number of secondary electrons generated on the radial surface of the electrode is not sufficient to sustain the discharge.

The same phenomenon occurs in our device at the ignition of the MHCD before reaching its steady state regime. Indeed, at the ignition, we observe a high current peak (Figure 5), which has about the same duration and peak amplitude as in the self-pulsing regime. In our normal working conditions, the ballast resistance is reduced to 39 kΩ, so that the time constant RC is about 2.4 µs. At the breakdown point, the voltage drops sharply and the current rises sharply in typically 1 µs. This behaviour is similar to an electrical switch connected with a resistance in series. Here, the equivalent resistance of the plasma varies during the formation of the plasma structure inside the MHCD cavity (sheath, positive column…). The typical current pulse duration is around 2 µs (as shown in Figure 5). Since the RC time constant is of the order of the discharging time, the plasma can be maintained in a stable regime, with a constant current value of 2.5mA for this particular ballast resistance.

### 4.1.2. Extinction phase

For the plasma extinction, a different behaviour was obtained. Figure 7 shows the current and voltage waveforms for the collapse phase of the microdischarges at (a) 350 and (b) 750 Torr He (zone B in Figure 4). The shown time windows start a few tens of micro seconds before the total extinction of the plasma inside the cavity. Combining the information in Figures 4 and 7(a), we can see that the current through the 350 Torr plasma decreases slowly from a few mA to roughly 60 µA before dropping quickly to zero.

The current appears to be slightly negative after 20 µs, but this is an experimental artefact: it is actually zero. At the same time, the voltage is found to increase slightly as the current decreases from 90 to roughly 60 µA in Figure 7(a) and rises quickly when the current drops. At this lower pressure (350 Torr), a monotonic and smooth decrease of the discharge current, resulting from an increase of the plasma impedance is observed. No oscillations form. This is a consequence of the high impedance of the plasma just before the extinction. Deduced from V/I in Fig. 7(a), the 1.6MΩ plasma impedance is much larger than the 39 kΩ ballast resistor and the characteristic RC value of 100 µs associated to 1.6MΩ is too large compared to the collapse time to permit the generation of oscillations under these conditions.

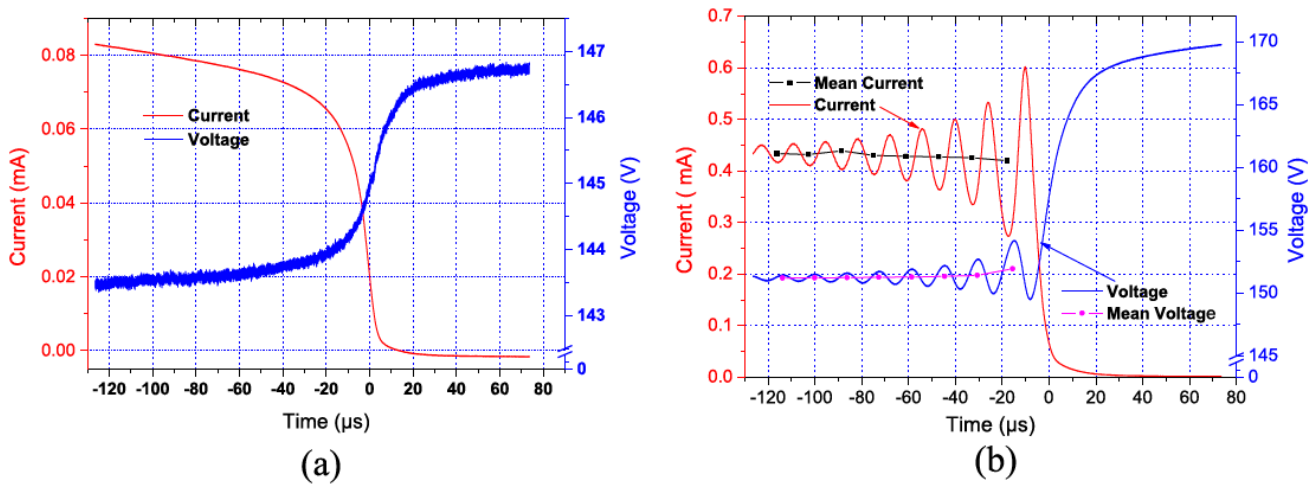

FIG. 7. Discharge current and voltage curves w. r. t. time for the collapse phase of the microdischarges at (a) 350 and (b) 750 Torr He.

Figure 7(b) shows the extinction of the MHCD at higher pressure (750 Torr). Here, it can be seen that, on both current and voltage, an oscillating regime starts a few tens of micro seconds before the complete extinction of the plasma. This amplified oscillatory behaviour of the current is completely different from the ending of the microdischarges at lower pressures (<400 Torr) and from the ignition phenomena. The period of the oscillation is around 14 µs, which should correspond to a characteristic oscillation period of the system. With the 360 kΩ impedance of the plasma, deduced from V/I in the beginning of Fig. 7(b), and the 60 pF capacitance of the microdischarge sandwich, the calculated RC would be 20 µs, slightly larger than the period of the observed oscillations. It should be remembered that this oscillating phenomenon is occurring in the decay period of the voltage supplied by the generator, during which the impedance of the plasma is continuously increasing. One can imagine that once the RC value approaches the natural oscillation period of the system, the impedance of the plasma starts to oscillate, ending with the total extinction of the plasma.

Figure 8 shows the averaged plasma impedances with respect to time for two different pressures 350 Torr and 750 Torr. From this figure, the change in impedance with pressure can be clearly observed. A current waveform of the transition from DC operation to self-pulsing mode just before plasma extinction at 400 Torr He is shown in Figure 9. This behaviour was not systematically obtained, but was occurring sometimes. One can observe at the transition of the amplified instability, which was usually preceding the extinction. But in this case, a self-pulsing phase was preceding the extinction.

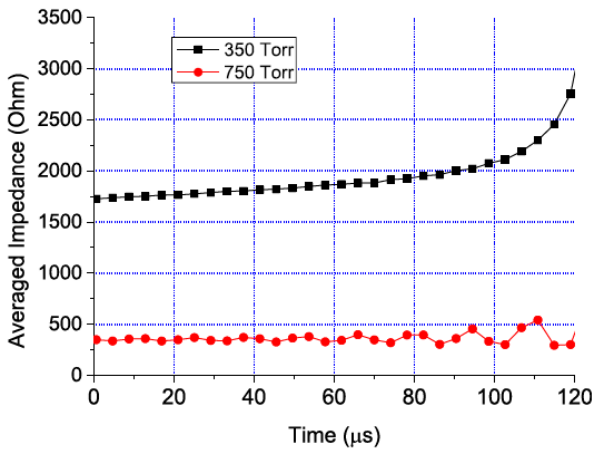

FIG. 8. Averaged plasma impedances vs. time for 350 and 750 Torr.

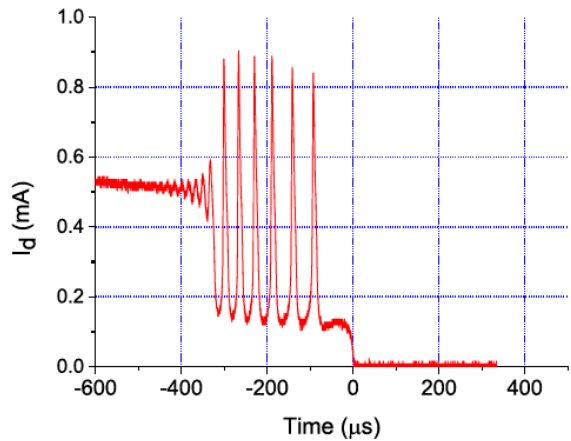

FIG. 9. Transition from DC operation to self-pulsing mode just before plasma extinction at 400 Torr He.

To make sure that these oscillations were not linked to an electrical noise, the emission signal, detected by a Photo Multiplier Tube (PMT), was recorded during the ending of the plasma. An example of such a PMT signal is shown in Figure 10. From this figure, clear oscillations in the light intensity of the microdischarge, in phase with the current oscillations, can be observed, which indicates that they are generated by the microdischarge itself. These PMT measurements are confirmed by the TDLAS experiments, which are presented in Sec. IV B.

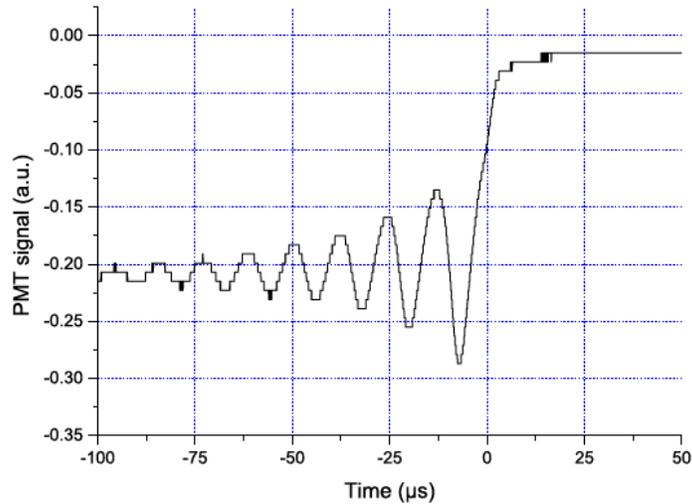

FIG. 10. PMT signal recorded for the extinction of the microdischarge.

## 4.2. TDLAS characterisation for ignition and extinction

To be able to measure the time varying density of metastable atoms in the ignition and the extinction of MHCD, we first evaluated the width of the absorption profiles for different discharge currents at two gas pressures: 350 and 750 Torr. Then, the laser frequency was set at the center of the line profile and the absorption signal was recorded during these periods. The MHCD was supplied with the triangular voltage ramp from the power supply. Assuming a Lorentzian profile, with a spectral width $\nu_L$ (FWHM) in unit of GHz, the density is related to the peak value of ln ($I_0/I$) at the center of the line profile, by the relation

$$N_i = \frac{\pi \Delta \nu_L}{2} \frac{1}{l f_{ik}} 3.8 \times 10^{14} \ln\left(\frac{I_0}{I}\right) = \frac{\Delta \nu_L}{l f_{ik}} 5.97 \times 10^{14} \ln\left(\frac{I_0}{I}\right)$$

where the oscillator strength to consider is $f_{ik} \approx 0.48$. Using this value, Eq. (1) becomes

$$N_i = \frac{\Delta \nu_L}{l} 1.24 \times 10^{15} \ln\left(\frac{I_0}{I}\right)$$

with the mean metastable density in m$^{-3}$ and l, the absorption length, in m. l=270 μm was used for the calculations. Note that the absorption length can be slightly changed if the pressure is varied due to the expansion of the plasma at the cathode side. As a consequence, the metastable can be a little bit overestimated, especially at low pressure.

Metastable densities were estimated for the starting and the ending of the plasma using Eq. (2). The value of $\Delta\nu_L$ was assumed to be identical to the one obtained from the absorption curves in stationary discharge (Figure 11) at the same pressure and the discharge current as following (-preceding-) the transient period in the ignition (-collapse-) of the plasma. This assumption will be discussed in Secs. IV B 1 and IV B 2 for each of ignition and collapse phases. Table I shows the used FWHM of the absorption profiles obtained under stable plasma operation at different pressures and discharge currents. Missing values have been deduced by interpolation of these data. Figure 12 is showing examples of the simulated experimental absorption profiles for two studied pressures (a) 350 Torr at 5mA and (b) 750 Torr at 10 mA) with best fit of corresponding Voigt functions. Although characteristic dimensions are short, the electric field is only high in the sheath and remains quite low in the main positive column of the discharge. Moreover, even though the electron density can reach a typical value of $10^{14}$ cm$^{-3}$, this value is not sufficient to provide such a significant broadening of the line.

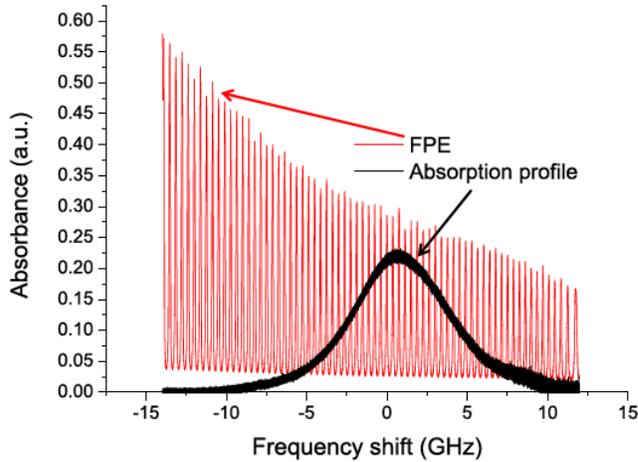

FIG. 11. Typical absorption profile obtained from TDLAS experiment under stable plasma operation at 500 Torr He and 15mA of discharge current.

| Pressure (Torr) | Current (mA) | FWHM (GHz) ± 0.3 |
|---|---|---|
| 350 Torr | 5 | 7.1 |
|  | 10 | 6.5 |
|  | 15 | 5.7 |
|  | 20 | 5.7 |
| 750 Torr | 5 | 14.0 |
|  | 10 | 9.1 |
|  | 15 | 7.7 |
|  | 20 | 8.7 |

TABLE I. FWHM of absorption profiles obtained under stable plasma operations, for different discharge currents at 350 and 750 Torr.

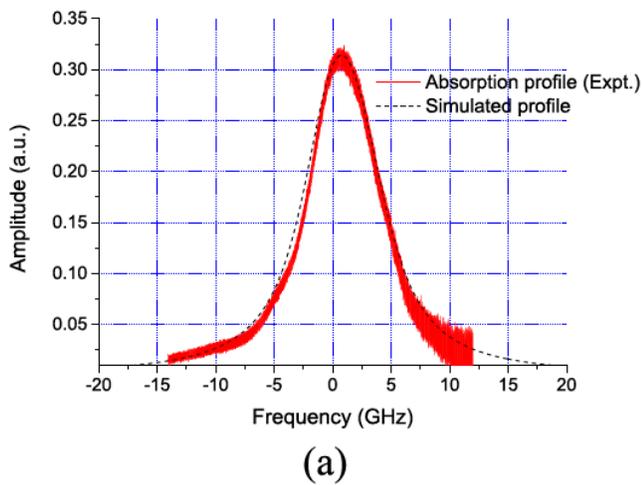
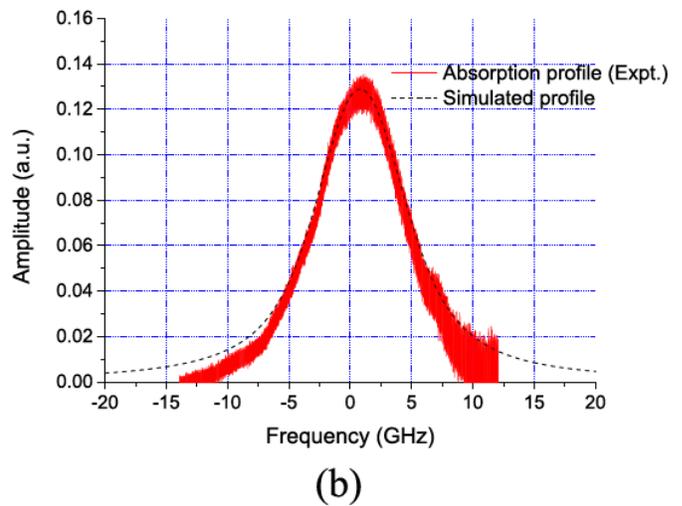

FIG. 12. Examples of the simulated experimental absorption profiles for two studied pressures (a) 350 Torr at 5mA and (b) 750 Torr at 10 mA with best fit of corresponding Voigt functions.

#### 4.2.1. Ignition

Figure 13(a) shows the time evolution of the metastable density and the discharge current during the ignition of the microplasma at 750 Torr in He. At the breakdown, the metastable density follows the rise of the current but after a small peak reaches a minimum at around 2 μs. One should first emphasize that due to the three-body quenching of He* metastable atoms

$$He^* + He + He \rightarrow He_2^* + He$$

The lifetime of these atoms at 750 Torr and room temperature, which is the gas temperature before the breakdown, is about 10 μs 20,21 and will even be longer at lower helium pressures. The much faster initial enhancement of the density reveals the presence of fast destruction mechanism by electrons, which could be quenching to the ground state,22 their stepwise ionization,20,23 or excitation to the singlet states followed by the radiative cascade to the ground state.22 After the initial small peak, the discharge current and different plasma parameters, as $n_e$, $T_e$, and $T_g$, will evolve toward their "stable discharge" values, reached at around 50 µs. The value of $\Delta \nu_L$ used to calculate the density of He* atoms reported in this figure corresponds to the one obtained in stable discharge at 750 Torr and 2.5 mA. As $n_e$, $T_e$, and $T_g$, should all slightly increase during this transient period, as shown in Fig. 13(a), the reported density for t<50 µs can be slightly underestimated, due to the under estimate of D L (overestimate of $T_g$). However, at 2.5 mA, the change in $\Delta \nu_L$ due to the gas heating will be less than 10% and within the measurements uncertainties. Afterwards, the discharge current stabilises at around 2.5 mA and the plasma switches to the normal regime, the metastable density remains constant at around $1.2*10^{18}$ m$^{-3}$. A similar behaviour was also observed at 350 Torr, with a metastable density of $7*10^{17}$ m$^{-3}$ for a discharge current of around 2 mA.

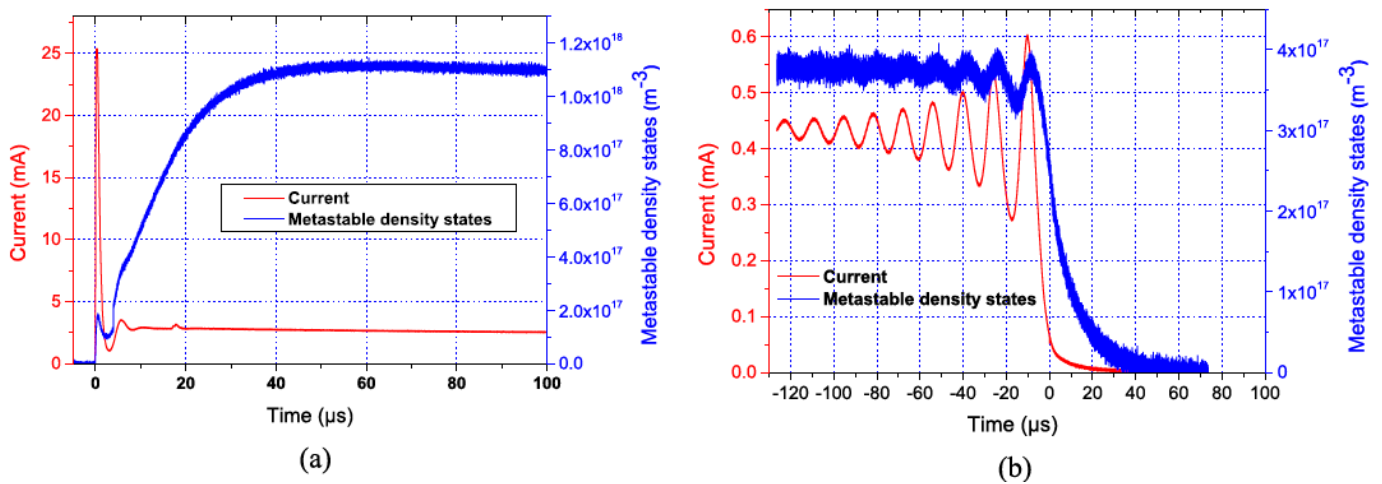

FIG. 13. Time evolution of metastable density and of discharge current at 750 Torr He in (a) starting (b) collapsing periods of the MHCD discharges run with slowly ramped DC voltage.

**4.2.2. Extinction**

Figure 13(b) shows the time evolution of the metastable density and the discharge current during the extinction of the microplasma at 750 Torr in He. Before the starting of the transient period of the collapse, the metastable density is around $4*10^{17}$ m$^{-3}$, at a discharge current of 0.45 mA. As in Fig. 7(b), the extinction of the plasma is preceded by an oscillatory regime of both current and metastable density, with an about 2 µs delay of the density relative to the current. This is a consequence of the finite lifetime of the metastable state, to which ground state atoms are excited, by electron impact, in phase with the discharge current. This delay attests that the density oscillation is a consequence of the current modulation, but not at its origin. The about 10 µs decay time of the metastable density following the falling of the current after its last maximum, corresponds to the lifetime of metastable atoms at 750 Torr and room temperature. In fact, at 0.45 mA, the gas temperature is very close to 300K and as, under these experimental conditions, the heat transport characteristic time inside the MHCD is about 50 µs and because during the current modulation, the mean current is about 0.45 mA, the gas temperature remains almost constant up to the total extinction of the plasma. Consequently, the value of D L to consider is always the same and the density evolution reported in Fig. 12(b) is correct. At pressures about and below 350 Torr, the current decreases to zero smoothly with the extinction of the plasma and hence the metastable density also smoothly decreases to zero from a value of $3*10^{17}$ m$^{-3}$ at 350 Torr. As it has already been discussed in Sec. IVA2, at low pressure, the resistance of the plasma before the oscillating regime starts is too high (as revealed by the less than 0.1mA discharge current in Fig. 7(a)) and the corresponding RC value is highly above the characteristic decay time of the current. Consequently, some oscillations will appear in the resistance of the plasma and the metastable density will follow the decay of the current, but more slowly because of its finite lifetime.

# 5. Conclusions

The ignition and the extinction phenomena for alumina based MHCD reactors were studied with a slowly ramped applied voltage. Helium was used for the experiments at pressures between 50 Torr and 1000 Torr. In the ignition phase of the plasma, it was found that the plasma started with a large current peak, which can reach tens of mA, depending on the gas pressure and the breakdown voltage. At the extinction of the plasma, for pressures higher than 400 Torr, oscillations appeared in both current and voltage curves when the ballast resistor was set below a certain value, depending on the gas pressure. It seemed that the period of these oscillations, with increasing amplitude up to the total extinction of the plasma, was a characteristic of the MHCD circuit and can be linked to the resistance of the plasma before reaching this transient regime. For the pressures below 400 Torr, the plasma extinguished with a smooth decrease of the current. The oscillating behaviors in the starting and collapsing periods of the plasma were attributed to the variations of the discharge resistance during these periods. To understand the origin of these oscillations and the eventual role played by helium metastable atoms in their appearance, TDLAS experiments were performed to measure the density of He*($^3S_1$) metastable atoms and to record the time variation of their density in different phases of the plasma and particularly during its ignition and extinction periods. It was found that the density of metastable varied according to current flowing through the MHCD plasma. At 750 Torr, the calculated metastable densities after the ignition and before the extinction of the microdischarges were found to be $1.2*10^{18}$ and $4*10^{17}$ m$^{-3}$, respectively. Metastable atoms did not seem to be at the origin of the generation of the observed instabilities, since the metastable density oscillations were behind current oscillations.